\documentclass[12pt, aps, prd, a4paper, amsmath, amssymb, groupedaddress, notitlepage, tightenlines, floatfix, nofootinbib, longbibliography]{revtex4-1}
\usepackage[T1]{fontenc}
\usepackage{graphicx}
\usepackage[pdftex, pdfstartview={FitH}, pdfnewwindow=true, colorlinks=true, citecolor=blue, filecolor=blue, linkcolor=blue, urlcolor=blue, pdfpagemode=UseNone]{hyperref}

\newcommand{\cD}{\ensuremath{\mathcal D} }
\newcommand{\ghat}{\ensuremath{\widehat{g}} }
\newcommand{\Gdag}{\ensuremath{G^{\dag}} }
\newcommand{\cN}{\ensuremath{\mathcal N} }
\newcommand{\cO}{\ensuremath{\mathcal O} }

\newcommand{\cQ}{\ensuremath{\mathcal Q} }
\newcommand{\Qbar}{\ensuremath{\overline Q} }
\newcommand{\Rbb}{\ensuremath{\mathbb R} }
\newcommand{\That}{\ensuremath{\widehat{T}} }
\newcommand{\cU}{\ensuremath{\mathcal U} }
\newcommand{\cUbar}{\ensuremath{\overline{\mathcal U}} }
\newcommand{\al}{\ensuremath{\alpha} }
\newcommand{\aldot}{\ensuremath{\dot\al} }
\newcommand{\be}{\ensuremath{\beta} }
\newcommand{\ga}{\ensuremath{\gamma} }
\newcommand{\gaEff}{\ensuremath{\ga_{\text{eff}}} }
\newcommand{\De}{\ensuremath{\Delta} }
\newcommand{\eps}{\ensuremath{\epsilon} }

\newcommand{\la}{\ensuremath{\lambda} }
\newcommand{\lalat}{\ensuremath{\la_{\text{lat}}} }
\newcommand{\muhat}{\ensuremath{\widehat{\mu}} }
\newcommand{\Om}{\ensuremath{\Omega} }

\newcommand{\si}{\ensuremath{\sigma} }
\newcommand{\SO}[1]{\ensuremath{\text{SO(}#1\text{)}} }
\newcommand{\SU}[1]{\ensuremath{\text{SU(}#1\text{)}} }
\newcommand{\U}[1]{\ensuremath{\text{U(}#1\text{)}} }
\newcommand{\vev}[1]{\ensuremath{\left\langle #1 \right\rangle} }
\newcommand{\gsim}{\ensuremath{\gtrsim} }

\newcommand{\llra}{\ensuremath{\longleftrightarrow} }
\newcommand{\X}{\ensuremath{\!\times\!} }
\newcommand{\pf}[0]{\ensuremath{\mbox{pf}\,} }
\newcommand{\Tr}[1]{\ensuremath{\mbox{Tr}\left[ #1 \right]} }

\newcommand{\fig}[1]{Fig.~\ref{#1}}
\newcommand{\secref}[1]{Sec.~\ref{#1}}
\newcommand{\refcite}[1]{Ref.~\cite{#1}}

\begin{document}
\setlength{\parskip}{10 pt}
\title{Lattice studies of supersymmetric gauge theories}
\author{David Schaich}\email{david.schaich@liverpool.ac.uk}
\affiliation{Department of Mathematical Sciences, University of Liverpool, Liverpool L69 7ZL, United Kingdom}
\date{17 August 2022}

\begin{abstract}
  Supersymmetry plays prominent roles in the study of quantum field theory and in many proposals for potential new physics beyond the standard model.
  Lattice field theory provides a non-perturbative regularization suitable for strongly interacting systems.
  This invited review briefly summarizes significant recent progress in lattice investigations of supersymmetric field theories, as well as some of the challenges that remain to be overcome.
  I focus on progress in three areas: supersymmetric Yang--Mills (SYM) theories in fewer than four space-time dimensions, as well as both minimal $\cN = 1$ SYM and maximal $\cN = 4$ SYM in four dimensions.
  I also highlight superQCD and sign problems as prominent challenges that will be important to address in future work.
\end{abstract}


\maketitle

\section{\label{sec:intro}Introduction} 
Supersymmetry plays several prominent roles in modern theoretical physics.
It is a valuable tool to improve our understanding of quantum field theory (QFT), an ingredient in many new physics models, and even a means to study quantum gravity via holographic duality.
Lattice field theory provides a non-perturbative regularization for QFTs, which has had enormous success as a means to analyze QCD and similar vector-like gauge theories.
It is therefore natural to explore how lattice field theory can be applied to investigate supersymmetric QFTs, especially in strongly coupled regimes.

In this short review I briefly summarize the recent progress and near-future prospects of lattice studies of supersymmetric systems.
This is an update and expansion of \refcite{Schaich:2018mmv}, incorporating pedagogical material presented at the 2021 online program ``Nonperturbative and Numerical Approaches to Quantum Gravity, String Theory and Holography'', run by the International Centre for Theoretical Sciences in Bengaluru.
To connect to the subject of holography featured by this Special Topics issue, I focus on four-dimensional gauge theories and their dimensional reductions to $d < 4$.
(See Refs.~\cite{Catterall:2009it, Kadoh:2016eju} for reviews of theories without gauge invariance, such as Wess--Zumino models and sigma models.  More recent work in this area includes Refs.~\cite{Kadoh:2019glu, Joseph:2020gdh, Dhindsa:2020ovr, Culver:2021rxo, Feng:2022nyi, Buividovich:2022jgv}.)

Lattice supersymmetry has been investigated for more than four decades~\cite{Dondi:1976tx}, and previously reviewed by Refs.~\cite{Catterall:2009it, Giedt:2009yd, Joseph:2015xwa, Kadoh:2016eju, Bergner:2016sbv, Hanada:2016jok, Schaich:2018mmv}, among others.
Unfortunately, progress in this area has been slower than for QCD-like theories, primarily because the lattice regularization of QFTs breaks supersymmetry.
This occurs in three main ways.
First, the super-Poincar\'e algebra includes the anti-commutation relation $\left\{Q_{\al}, \Qbar_{\aldot}\right\} = 2\si_{\al\aldot}^{\mu} P_{\mu}$ that connects the spinorial generators of supersymmetry transformations, $Q_{\al}$ and $\Qbar_{\aldot}$, to the generator of infinitesimal space-time translations, $P_{\mu}$.
Lattice regularization formulates the QFT of interest in a discrete space-time where no such infinitesimal translations exist, implying broken supersymmetry.

Second, bosonic and fermionic fields are typically discretized differently on the lattice.
In the context of supersymmetric gauge theories, standard discretizations associate the gauginos $\la_{\al}(n)$ with lattice sites while the gauge connections are associated with links $U_{\mu}(n)$ between nearest-neighbor sites.
That is, under a lattice gauge transformation $\la_{\al}(n) \to G(n) \la_{\al}(n) \Gdag(n)$ while $U_{\mu}(n) \to G(n) U_{\mu}(n) \Gdag(n + a\muhat)$, where `$a$' is the lattice spacing.
Away from the $a \to 0$ continuum limit, these differences prevent supersymmetry transformations from correctly interchanging superpartners.
Although scalar fields also tend to be associated with lattice sites, their discretization typically omits the features (e.g., a Wilson term or staggering) required to address the famous fermion doubling problem, resulting in a similar breaking of supersymmetry.

Finally, the Leibniz rule $\partial \left[\phi \eta\right] = \left[\partial \phi\right]\eta + \phi \partial \eta$ plays an important role in ensuring supersymmetry, but is violated by standard lattice finite-difference operators~\cite{Dondi:1976tx}.
In discrete space-time, `no-go theorems' presented by Refs.~\cite{Kato:2008sp, Bergner:2009vg} establish that only non-local derivative and product operators can obey the Leibniz rule and hence fully preserve supersymmetry.
This implies a trade-off between locality and supersymmetry, which continues to be explored for simple systems such as supersymmetric quantum mechanics, where a lattice field product operator obeying a `cyclic Leibniz rule' suffices to preserve partial supersymmetry and establish non-renormalization~\cite{Kato:2013sba, Kato:2016fpg, Kato:2018kop, Kadoh:2019bir}.
A different non-local `star product' is able to satisfy the Leibniz rule, but in such a way that the lattice spacing no longer acts as a regulator~\cite{DAdda:2017bzo}.
Even restricted to simple systems without gauge invariance, and mostly in 0+1~dimensions, these constructions already become very complicated, and the remainder of this review will focus on approaches that preserve locality at the expense of broken supersymmetry.

The breaking of supersymmetry in lattice calculations has the consequence that quantum effects generate supersymmetry-violating operators.
Of particular concern are \textit{relevant} supersymmetry-violating operators, for which counterterms have to be fine-tuned in order to recover the supersymmetric QFT of interest in the $a \to 0$ continuum limit that corresponds to removing the UV cutoff $a^{-1}$.
Considering the case of four space-time dimensions, many relevant operators can appear for theories that involve scalar fields.
Such theories include supersymmetric QCD (superQCD) with scalar squarks, as well as gauge theories with `$\cN > 1$' extended supersymmetry, which include scalar fields in the gauge supermultiplet.
The mass terms of these scalars introduce fine-tuning problems similar to that of the standard model Higgs, and additional relevant supersymmetry-violating operators can arise from the fermion (quark and gaugino) mass terms, Yukawa couplings, and quartic (four-scalar) terms.
Careful counting typically finds $\cO(10)$ relevant operators for lattice discretizations of supersymmetric theories with scalar fields~\cite{Giedt:2009yd, Elliott:2008jp, Catterall:2014mha}.
Simultaneously fine-tuning counterterms for all of these operators in numerical lattice calculations appears impractical, to say the least.

In order to make lattice supersymmetry practical, the amount of fine-tuning needs to be reduced.
The following three sections briefly review three different ways to achieve this for lattice studies of supersymmetric Yang--Mills (SYM) theories.
First, the next section discusses dimensional reductions of SYM theories to fewer than four space-time dimensions, which has been the focus of a great deal of recent work.
Returning to four dimensions, \secref{sec:min} considers the special case of minimal ($\cN = 1$) SYM, which is significantly simplified by the absence of scalar fields.
Maximal ($\cN = 4$) SYM is another special case in four dimensions, for which fine-tuning can be vastly reduced by preserving a closed subalgebra of the supersymmetries, as discussed in \secref{sec:max}.
We conclude in \secref{sec:future} by highlighting some prominent challenges to be faced by future lattice studies of supersymmetric QFTs, including superQCD and sign problems that can arise in several contexts.

\section{\label{sec:lowd}Dimensionally reduced SYM theories} 
Working in a smaller number of space-time dimensions, $d < 4$, can make numerical analyses much more tractable.
In addition to the smaller number of degrees of freedom corresponding to $L^d$ lattices, lower-dimensional theories tend to be super-renormalizable and in many cases a one-loop counterterm calculation suffices to restore supersymmetry in the continuum limit~\cite{Giedt:2004vb, Bergner:2007pu, Giedt:2018ygt}.
Focusing here on dimensional reductions of SYM theories, we will label systems by the number $Q$ of supersymmetry generators, or `supercharges', that they have.
Starting in four dimensions, $\cN = 1$, $2$ or $4$ SYM corresponds to $Q = 4$, $8$ or $16$, respectively.
These theories can also be considered dimensional reductions of minimal SYM in respectively $D_* = 4$, $6$ or $10$ dimensions.
For $d \leq 4$, these theories involve a $d$-component gauge field, fermionic fields with $Q$ total components, and $D_* - d$ real scalar fields.
All fields are massless and transform in the adjoint representation of the gauge group, here taken to be either SU($N$) or $\U{N} = \SU{N} \otimes \U{1}$.

\subsection{\label{sec:QM}0+1 dimensions} 
Dimensional reduction all the way to (0+1)-dimensional `SYM quantum mechanics' (QM) has been the subject of many numerical studies over the past fifteen years, starting with Refs.~\cite{Hanada:2007ti, Catterall:2007fp} investigating the $Q = 4$ case and also including Refs~\cite{Anagnostopoulos:2007fw, Hanada:2008gy, Hanada:2008ez, Catterall:2008yz, Hanada:2009ne, Catterall:2009xn, Catterall:2010gf, Hanada:2011fq, Hanada:2013rga, Honda:2013nfa, Steinhauer:2014oda, Ambrozinski:2014oka, Kadoh:2015mka, Filev:2015hia, Bergner:2016qbz, Hanada:2016zxj, Berkowitz:2016jlq, Rinaldi:2017mjl, Buividovich:2018scl, Berkowitz:2018qhn, Asano:2018nol, Rinaldi:2021jbg, Bergner:2021goh, Schaich:2022duk, Pateloudis:2022oos}.
These SYM QM systems consist of balanced collections of interacting bosonic and fermionic $N\X N$ matrices evolving in (euclidean) time at a single spatial point.
They are simple enough that lattice regularization may not even be required to analyze them --- Refs.~\cite{Hanada:2007ti, Anagnostopoulos:2007fw, Hanada:2008gy, Hanada:2008ez, Hanada:2009ne, Hanada:2011fq, Hanada:2013rga} instead employ a gauge-fixed Monte Carlo approach with a hard momentum cutoff, \refcite{Buividovich:2022jgv} exactly diagonalizes truncated hamiltonians, and Refs.~\cite{Buser:2020cvn, Gharibyan:2020bab, Rinaldi:2021jbg, Culver:2021rxo, Feng:2022nyi} explore prospects for quantum computing.
Another aspect of this simplicity is the proposal that $Q = 16$ SYM QM can be `ungauged' to produce a scalar--fermion system with SU($N$) global symmetry, with both the gauged and ungauged models flowing to the same theory in the IR~\cite{Maldacena:2018vsr, Berkowitz:2018qhn, Pateloudis:2022oos}. 

Even though SYM QM systems are much simpler to study on the lattice than their four-dimensional SYM counterparts, they remain computationally non-trivial.
Let's consider this in the context of the maximally supersymmetric $Q = 16$ case, which has attracted particular interest due to its connections to string theory~\cite{deWit:1988wri} --- especially the conjecture by \refcite{Banks:1996vh} that the large-$N$ limit of this system describes the strong-coupling (`M-theory') limit of type-IIA string theory in light-front coordinates.
Another contribution to this Special Topics issue will review this subject in more detail, and earlier reviews from string theory perspectives include Refs.~\cite{Taylor:1999qk, Ydri:2017ncg}.
At finite temperature, this holographic conjecture relates the bosonic action of the deconfined SYM QM system to the internal energy of a dual compactified black hole geometry in eleven-dimensional M-theory.

This quantity is straightforward to compute through numerical Monte Carlo analyses~\cite{Anagnostopoulos:2007fw, Hanada:2008ez, Hanada:2013rga, Catterall:2008yz, Catterall:2009xn, Kadoh:2015mka, Filev:2015hia, Hanada:2016zxj, Berkowitz:2016jlq}, with \refcite{Berkowitz:2016jlq} representing the state of the art that improves upon earlier results by carrying out controlled extrapolations to the large-$N$ continuum limit.
In addition to the role of large $N$ in holography, the absence of any spatial volume means that in these studies the thermodynamic limit itself corresponds to extrapolating $N \to \infty$.
These controlled extrapolations enable more robust comparisons to dual gravitational predictions, with numerical results convincingly approaching the leading-order gravitational prediction from classical supergravity at low temperatures, providing non-perturbative first-principles evidence that the holographic duality conjecture is correct.
In addition, deviations between the lattice results and leading-order supergravity at higher temperatures can be considered a prediction of higher-order quantum gravitational effects that are enormously difficult to calculate analytically.

The main computational challenge of these SYM QM investigations comes from the large numbers of colors $N$ and lattice sizes $L$ that are needed to control the large-$N$ continuum extrapolations.
In this context, with a fixed dimensionless temperature $\That \equiv T / \la^{1 / 3}$, the continuum limit corresponds to $L \to \infty$.
Here the $~\widehat{\cdot}~$ decoration highlights dimensionless ratios that can be considered consistently in both the lattice and continuum theories --- note the 't~Hooft coupling $\la = g_{\text{YM}}^2 N$ has dimension $[\la] = 4 - d$ in $d$ dimensions.
\refcite{Berkowitz:2016jlq} employs $16 \leq N \leq 32$ and lattice sizes up to $L = 64$, with the $\cO\!\left(N^3\right)$ cost scaling of matrix--matrix multiplication dominating over the $\sim$$L^{5d / 4}$ cost scaling of the rational hybrid Monte Carlo (RHMC) algorithm, and requiring large-scale supercomputing.

Large values of $N$ are also motivated by a thermal instability associated with the non-compact quantum moduli space of $Q = 16$ SYM QM~\cite{Catterall:2009xn}.
At low temperatures the system is able to run away along these flat directions, which is interpreted holographically as D$0$-brane radiation from the dual black hole.
A formal solution is to stabilize the desired vacuum by adding a supersymmetry-breaking scalar potential to the lattice action, which then needs to be removed in the course of extrapolating to the continuum limit~\cite{Catterall:2009xn, Hanada:2009hq}.
To avoid this complication, \refcite{Berkowitz:2016jlq} argues that in practice it can be possible to carry out Monte Carlo sampling around a metastable vacuum so long as $N$ is large enough to reduce the tunneling rate to the true run-away vacuum.
The necessary value of $N$ increases as the temperature decreases.

An alternative is to consider a deformation of $Q = 16$ SYM QM introduced by Berenstein, Maldacena and Nastase (BMN)~\cite{Berenstein:2002jq}, which lifts the flat directions mentioned above while preserving all $16$ supercharges.
This deformation serves as a supersymmetric regulator that doesn't need to be removed in the continuum limit.
It introduces non-zero masses for the $9$ scalars and $16$ fermions of the theory, explicitly breaking the SO(9) global symmetry associated with the compactified spatial dimensions of $d = 10$ minimal SYM, $\SO{9} \to \SO{6}\X \SO{3}$.
The deformation depends on a dimensionful mass parameter $\mu$, which can be combined with the 't~Hooft coupling to define a dimensionless coupling $\ghat \equiv \la / \mu^3$ (not to be confused with the dimensionful Yang--Mills gauge coupling $g_{\text{YM}}^2 = \la / N$).

This BMN model has been studied numerically by Refs.~\cite{Catterall:2010gf, Honda:2013nfa, Asano:2018nol, Rinaldi:2021jbg, Bergner:2021goh, Schaich:2022duk, Pateloudis:2022oos}, with particular focus on its finite-temperature confinement transition.
The critical temperature $\That_c$ of this transition can be predicted both by perturbative calculations in the weak-coupling regime $\ghat \ll 1$~\cite{Furuuchi:2003sy, Spradlin:2004sx, Hadizadeh:2004bf} as well as by dual supergravity calculations for strong couplings $\ghat \to \infty$ with $N \to \infty$ and $\That \ll 1$~\cite{Costa:2014wya}.
The goals of ongoing lattice calculations include both reproducing these limits as well as non-perturbatively connecting them by mapping out the intermediate regime where perturbative and holographic approaches are unreliable.
Good progress has been made by Refs.~\cite{Asano:2018nol, Bergner:2021goh, Schaich:2022duk}, considering different lattice discretizations, numbers of colors $N$, lattice sizes $L$, and couplings \ghat that span several orders of magnitude.

\subsection{\label{sec:2d}1+1 dimensions} 
Moving from (0+1)-dimensional quantum mechanics to QFTs in two dimensions introduces many additional phenomena to explore, while remaining significantly more tractable than numerical studies in four dimensions.
Research on supersymmetric lattice gauge theories in $2 \leq d \leq 4$ dimensions features both conceptual work constructing clever lattice formulations that minimize fine-tuning in principle, as well as numerical work that exploits these constructions to carry out practical calculations.
The most widely applied reformulations are based on approaches known as topological twisting~\cite{Sugino:2003yb, Sugino:2004qd, Catterall:2004np} and orbifolded dimensional deconstruction~\cite{Cohen:2003xe, Cohen:2003qw, Kaplan:2005ta, Unsal:2006qp}, which are thoroughly reviewed by \refcite{Catterall:2009it}.
While the concepts and terminology of these approaches differ, in the end they actually produce equivalent constructions~\cite{Catterall:2007kn, Damgaard:2008pa}.

Here we will only briefly summarize the twisting approach.
Restricting our considerations to flat space-time, twisting is just a change of variables that organizes linear combinations of the supercharges into completely antisymmetric $p$-forms $\cQ$, $\cQ_{\mu}$, $\cQ_{\mu\nu}$, etc., with the feature that any twisted-scalar supercharge is nilpotent, $\cQ^2 = 0$.
This is only possible if there are at least $2^d$ supercharges in $d$ dimensions, and provides at most $\lfloor Q / 2^d \rfloor$ nilpotent $\cQ$.
The $p$-forms transform with integer `spin' under the so-called twisted rotation group $\SO{d}_{\text{tw}} \equiv \mbox{diag}\left[\SO{d}_{\textrm{euc}} \otimes \SO{d}_R\right]$, where $\SO{d}_{\textrm{euc}}$ is the Wick-rotated Lorentz group and $\SO{d}_R$ comes from the global $R$-symmetry.
This procedure clearly provides a closed supersymmetry subalgebra $\left\{\cQ, \cQ\right\} = 0$ that can be preserved at non-zero lattice spacing, and ultimately leads to a $\cQ$-invariant lattice action with no need of the Leibniz rule.

Consistently with the discussion in \secref{sec:intro}, the other twisted supercharges $\cQ_{\mu}$, $\cQ_{\mu\nu}$, $\cdots$, are all broken by the lattice discretization.
However, the preservation of a subset of the supersymmetries and a closed subalgebra of the full super-Poincar\'e algebra vastly reduces the fine-tuning required to recover the other, broken supersymmetries in the continuum limit.
We will comment on this in more detail for four-dimensional $\cN = 4$ SYM in \secref{sec:max}, but it is useful here to consider an analogy with the more basic Poincar\'e symmetries of translations and rotations.
While these continuous symmetries are also broken in discrete lattice space-times, since the early days of lattice field theory it has been appreciated that the discrete symmetries preserved by (e.g.) hypercubic lattices guarantee their recovery in the continuum limit, with no need for fine-tuning.
In a similar way, the preserved twisted-scalar supersymmetries \cQ guarantee the recovery of the full set of supersymmetries in the continuum limit, with little to no fine-tuning.

For some theories there are multiple, inequivalent ways that twisted lattice systems can be constructed.
One approach~\cite{Catterall:2007kn, Catterall:2014vka, Schaich:2014pda, Catterall:2015ira} combines the gauge and scalar fields into a complexified gauge field.
This results in $\U{N} = \SU{N} \otimes \U{1}$ gauge invariance and non-compact lattice gauge links $\left\{\cU, \cUbar\right\}$ with a flat measure.
Although the $\U{1}$ sector decouples in the continuum, at non-zero lattice spacing it can introduce unwanted artifacts, especially at strong couplings.
Refs.~\cite{Catterall:2015ira, Catterall:2017lub, Catterall:2018laz, Catterall:2020lsi} explore various ways these artifacts could be suppressed for four-dimensional $\cN = 4$ SYM, which we will revisit in \secref{sec:max}.
Numerical studies using this formulation for $Q = 16$ SYM in two dimensions include Refs.~\cite{Catterall:2008dv, Catterall:2010fx, Catterall:2011aa, Catterall:2017lub}.

A different approach~\cite{Sugino:2003yb, Sugino:2004qd, Kadoh:2009rw, Hanada:2010kt, Hanada:2011qx, Matsuura:2014kha, Hanada:2017gqc} works with compact gauge links and gauge group SU($N$), at the cost of imposing an admissibility condition to resolve a huge degeneracy of vacua.
While \refcite{Matsuura:2014pua} proposes formulations of $Q = 4$ and $Q = 8$ SYM in two dimensions that avoid the need to impose admissibility conditions, this issue becomes more problematic in higher dimensions~\cite{Sugino:2004uv, Catterall:2009it}.
Numerical studies using this formulation in two dimensions include Refs.~\cite{Hanada:2009hq, Suzuki:2007jt, Kanamori:2007ye, Kanamori:2007yx, Kanamori:2008bk, Kanamori:2008yy, Kanamori:2009dk, Hanada:2010qg, Kamata:2016xmu, Ohta:2021jze} investigating $Q = 4$ SYM and Refs.~\cite{Giguere:2015cga, Kadoh:2017mcj} investigating $Q = 16$ SYM.

\begin{figure*}[tbp]
  \includegraphics[height=0.35\linewidth]{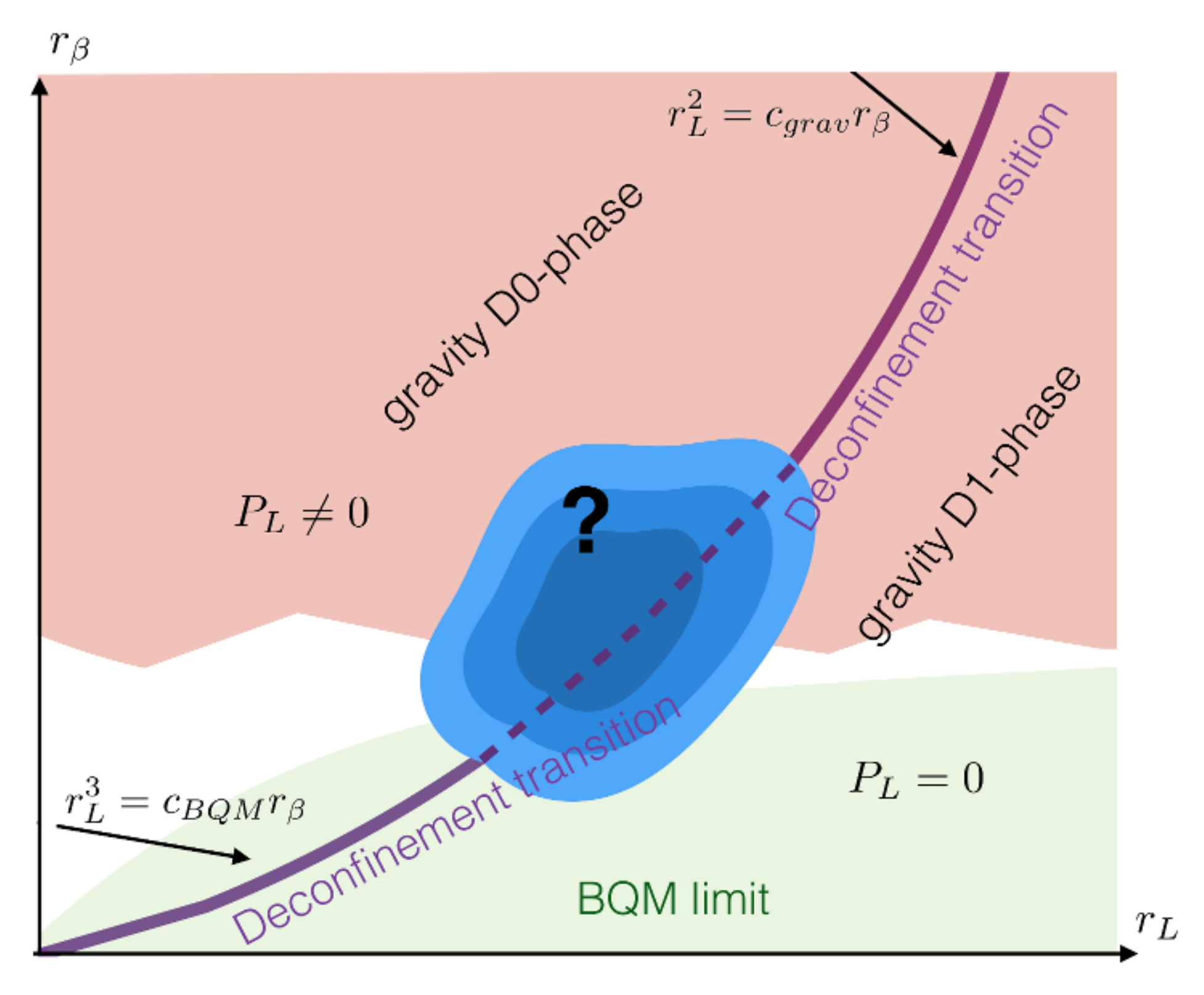}\hfill \includegraphics[height=0.35\linewidth]{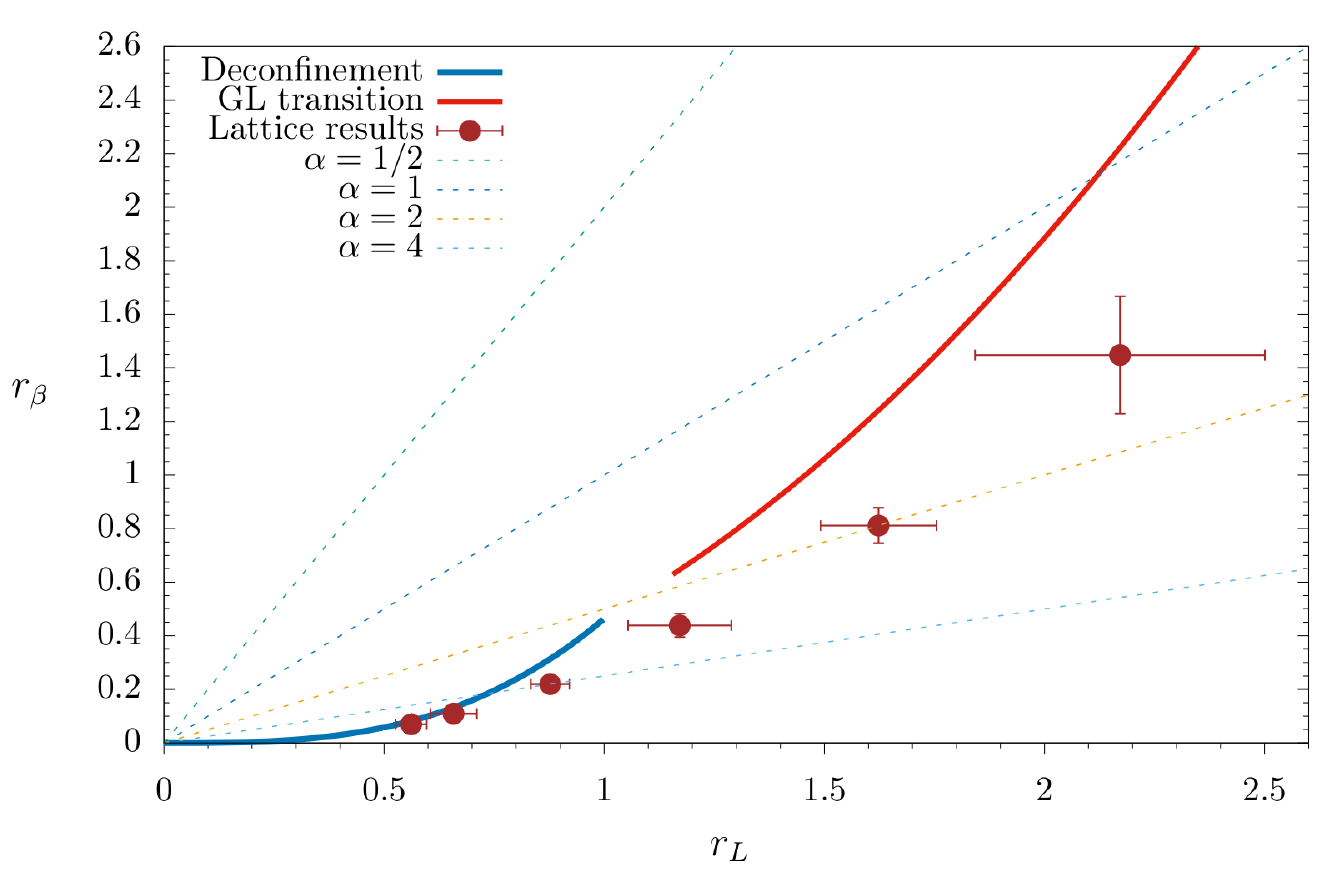}
  \caption{\label{fig:2dQ16}A sketch (\textbf{left}) of the expected phase diagram for two-dimensional $Q = 16$ SYM, compared to numerical results (\textbf{right}), both adapted from \refcite{Catterall:2017lub}.  The numerical calculations use a twisted formulation with gauge group SU($12$) and aspect ratios $\al = r_L / r_{\be} = N_s / N_t$ ranging from $8$ to $3 / 2$.}
\end{figure*}

As for the BMN model discussed in \secref{sec:QM}, a prominent physics target is to map out the non-perturbative phase diagrams of two-dimensional SYM theories~\cite{Catterall:2017lub, Jha:2018acq, Dhindsa:2021irw}.
These systems are now formulated on an $r_L \X r_{\be}$ torus, where $r_{\be} = 1 / \That$ is the inverse dimensionless temperature while $r_L = L \sqrt{\la}$ is the dimensionless length of the spatial cycle.
At high temperatures corresponding to small $r_{\be}$, the fermions pick up a large thermal mass, reducing the system to bosonic quantum mechanics (BQM) in one dimension.
This is illustrated for $Q = 16$ SYM by the sketch in the left panel of \fig{fig:2dQ16}, which highlights the `spatial deconfinement' transition expected as $r_L$ decreases in the large-$N$ limit.
(The system is always thermally deconfined.)
In the BQM limit, this transition appears to be first order --- see \refcite{Bergner:2019rca} and references therein.

A similar first-order transition is predicted by holography in the large-$N$, low-temperature regime corresponding to large $r_{\be}$.
Holographically, the large-$r_L$ spatially confined phase is conjectured to be dual to a homogeneous black string with a horizon wrapping around the spatial cycle, while the small-$r_L$ spatially deconfined phase corresponds to a localized black hole.
Challenging numerical supergravity calculations are required to construct and analyze these dual black hole and black string geometries~\cite{Dias:2017uyv}.

The right panel of \fig{fig:2dQ16} presents lattice results for the $Q = 16$ SYM spatial deconfinement transition from \refcite{Catterall:2017lub}, again aiming to reproduce the expected high- and low-temperature limits while non-perturbatively mapping out intermediate temperatures.
To this end the calculations vary $r_L$ with a fixed aspect ratio $\al = r_L / r_{\be} = N_s / N_t$ defined by the $N_s\X N_t$ lattice size, monitoring the spatial Wilson line $\Tr{\prod_{x_i} \cU_x(x_i, t)}$ as the order parameter for this transition.
As shown by the dashed lines in the plot, larger aspect ratios (up to $\al = 8$ for a $32\X 4$ lattice) probe the spatial deconfinement transition at higher temperatures, matching the BQM expectations quite well.
As the aspect ratio decreases (down to $\al = 3 / 2$ for an $18\X 12$ lattice), the lower-temperature numerical results are consistent with holography, albeit with rapidly increasing uncertainties.
At these low temperatures, a supersymmetry-breaking scalar potential is added to the lattice action to lift flat directions, and then extrapolated to zero, as discussed in \secref{sec:QM}.
\refcite{Catterall:2017lub} also calculates the internal energies of the dual black hole and black string, finding consistency with holographic expectations in both phases, again with large uncertainties.
Although the work compares multiple lattice volumes and SU($N$) gauge groups up to $N = 16$, room for improvement remains in terms of carrying out controlled extrapolations to the continuum and thermodynamic (i.e., large-$N$) limits.
In particular, larger values of $N$ should help to reduce uncertainties and access even lower temperatures.

Beyond investigations of phase diagrams, two-dimensional SYM theories also possess rich zero-temperature dynamics that are important to explore non-perturbatively.
For example, \refcite{August:2018esp} analyzes the `meson' spectrum of the $Q = 4$ lattice theory, using a straightforward Wilson-fermion discretization rather than a twisted construction, and observing a massless supermultiplet predicted by Refs.~\cite{Witten:1995im, Fukaya:2006mg}.
This work also checks for spontaneous supersymmetry breaking (SSB), which \refcite{Hori:2006dk} suggests might occur for this theory.
While SSB is also being explored for supersymmetric QM in 0+1~dimensions~\cite{Baumgartner:2014nka, Baumgartner:2015qba, Baumgartner:2015zna, Joseph:2020gdh, Dhindsa:2020ovr, Culver:2021rxo}, this can be a truly dynamical process in two dimensions~\cite{Wozar:2011gu, Steinhauer:2014yaa}, rather than being determined by the superpotential.
We will revisit SSB in the context of the sign problem in \secref{sec:sign}, for now simply noting that \refcite{August:2018esp} saw no evidence of SSB for $Q = 4$ SYM, consistent with other lattice studies using twisted formulations~\cite{Kanamori:2007ye, Kanamori:2007yx, Kanamori:2009dk, Catterall:2017xox}.

\subsection{2+1 dimensions} 
\begin{figure*}[tbp]
  \includegraphics[height=0.34\linewidth]{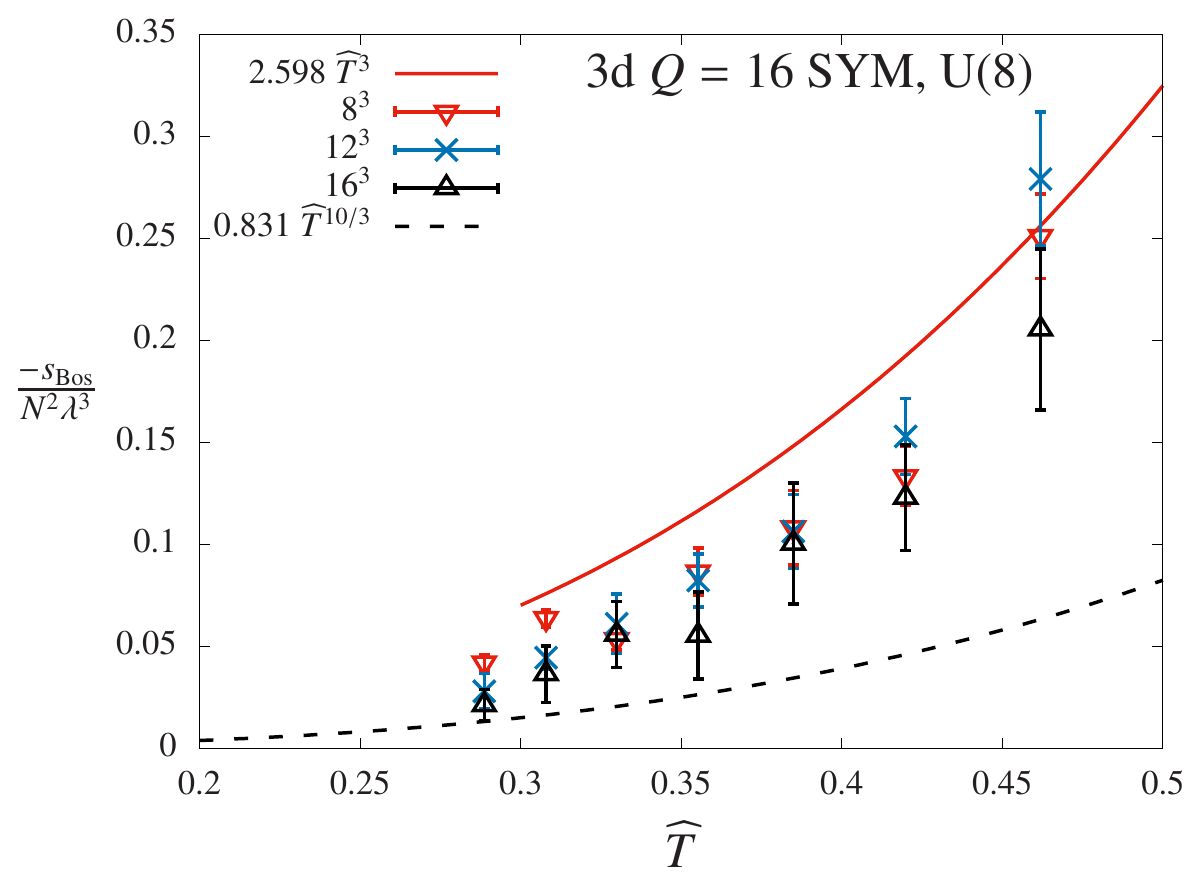}\hfill \includegraphics[height=0.34\linewidth]{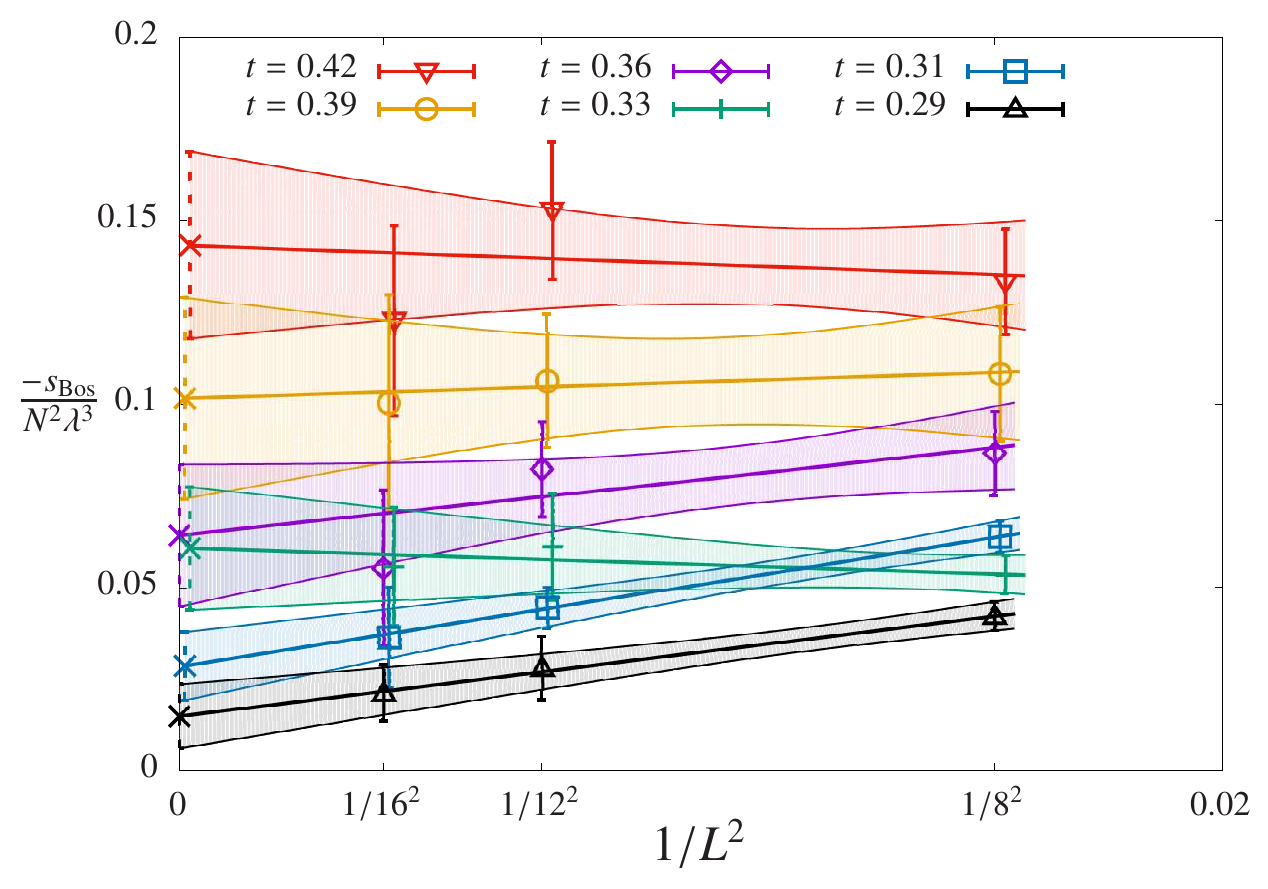}
  \caption{\label{fig:3dQ16}Bosonic action density for three-dimensional $Q = 16$ SYM with gauge group U($8$) and $L^3$ lattice volumes, from \refcite{Sherletov:2022rnl}.  \textbf{Left:} Results for $L = 8$, $12$ and $16$ vs.\ the dimensionless temperature $\That$, compared with the high-temperature expectation $\propto$$\That^3$ and the low-temperature dual-supergravity prediction $\propto$$\That^{10 / 3}$.  \textbf{Right:} Linear $L^2 \to \infty$ continuum extrapolations for the six lowest temperatures.}
\end{figure*}

Three-dimensional SYM remains a prominent frontier for lattice studies, with many compelling physics targets and more modest computational costs compared to $d = 4$.
Supersymmetry-preserving twisted formulations (discussed by Refs.~\cite{Catterall:2011cea, Giedt:2017fck, Giedt:2018ygt}) are even more important than in two dimensions, and are used by all numerical calculations so far~\cite{Catterall:2020nmn, Sherletov:2022rnl}.
Figure~\ref{fig:3dQ16} presents some results from these works, investigating the $Q = 16$ SYM bosonic action that holography relates to the internal energy of the dual black brane geometry in supergravity.
Here the calculations use $L^3$ lattice volumes corresponding to an aspect ratio $\al = 1$, and are kept within the spatially confined phase dual to a homogeneous black D$2$-brane.
The results in the left panel of \fig{fig:3dQ16} approach the corresponding supergravity prediction for sufficiently low dimensionless temperatures.
While these calculations consider gauge group U($N$) with only a single $N = 8$, one notable advance are the extrapolations to the continuum limit shown in the right panel of \fig{fig:3dQ16}, which were not attempted in the $d = 2$ study that produced \fig{fig:2dQ16}.
As in lower dimensions, with a fixed dimensionless temperature and $\al = 1$, the continuum limit corresponds to extrapolating $L \to \infty$ while the thermodynamic limit would be provided by extrapolating $N \to \infty$.

Ongoing work is now building on these results to explore the phase transition between this `D$2$ phase' and the spatially deconfined `D$0$ phase' dual to a localized black hole geometry.
This transition corresponds to considering an $r_L \X r_L \X r_{\be}$ three-torus, with other behavior possible if the two spatial cycles are allowed to have different sizes~\cite{Morita:2014ypa}.
\refcite{Sherletov:2022rnl} also presents initial investigations of $Q = 8$ SYM in three dimensions, for which new code is being developed within the publicly available \texttt{SUSY LATTICE} parallel software package for twisted lattice supersymmetry~\cite{Schaich:2014pda, Bergner:2021ffz_code}. 

\section{\label{sec:min}Minimal $\cN = 1$ SYM in four dimensions} 
In \secref{sec:intro} we emphasized that scalar fields are responsible for most of the relevant supersymmetry-violating operators that require fine-tuning in lattice calculations, including the scalar mass terms, Yukawa couplings, and quartic operators.
This implies that upon reaching $d = 4$ dimensions, the most promising supersymmetric gauge theory to consider would be $\cN = 1$ SYM, the only one with no scalar fields.
$\cN = 1$ SYM involves only the SU($N$) gauge field and its superpartner gaugino --- a massless Majorana fermion transforming in the adjoint rep of SU($N$).
The mass term for this gaugino is the only relevant operator that may need to be fine-tuned in order to obtain the correct continuum limit~\cite{Curci:1986sm, Suzuki:2012pc}.
In fact, even this single fine-tuning can be avoided by using overlap or domain-wall lattice fermion formulations that obey the Ginsparg--Wilson relation and preserve chiral symmetry at non-zero lattice spacing.
Even though the axial anomaly breaks the classical $\U{1}$ $R$-symmetry of $\cN = 1$ SYM to its $Z_{2N}$ subgroup, preserving this discrete global symmetry still protects the gaugino mass from additive renormalization.
The spontaneous breaking of $Z_{2N} \to Z_2$ by the formation of a gaugino condensate, $\vev{\la\la} \ne 0$, is also relatively straightforward to investigate with this approach~\cite{Giedt:2008xm, Endres:2009yp, Kim:2011fw, Piemonte:2020wkm}.

The downside to overlap and domain-wall fermions is that they are computationally expensive, and after Refs.~\cite{Giedt:2008xm, Endres:2009yp, Kim:2011fw} there has been very little work applying them to $\cN = 1$ SYM during the past decade~\cite{Piemonte:2020wkm}.
Instead, recent lattice research into this theory has employed improved Wilson fermions, fine-tuning the gaugino mass to recover both chiral symmetry and supersymmetry in the continuum limit.
For example, the DESY--M{\"u}nster--Regensburg--Jena Collaboration made significant progress using clover-improved Wilson fermions~\cite{Bergner:2014saa, Bergner:2014dua, Bergner:2014ska, Ali:2018dnd, Ali:2018fbq, Bergner:2018unx, Ali:2019gzj, Ali:2019agk, Ali:2020mvj, Ali:2021zzk}.
A second group is exploring a SYM analogue of the twisted-mass fermion action~\cite{Steinhauser:2020zth}, with the aim of improving the formation of composite supermultiplets at non-zero gaugino masses and lattice spacings, and thereby gaining better control over the chiral and continuum extrapolations.

With limited connections to holography, and a spatial lattice volume to provide the thermodynamic limit, there is little motivation for extremely large values of $N$, and most work considers only gauge groups SU($2$)~\cite{Bergner:2014saa, Bergner:2014dua, Bergner:2014ska, Ali:2019gzj, Ali:2021zzk} and SU($3$)~\cite{Ali:2018dnd, Ali:2018fbq, Bergner:2018unx, Ali:2019agk, Ali:2020mvj, Steinhauser:2020zth}.
A long-term goal of these studies has been to observe the composite states forming degenerate multiplets in the supersymmetric continuum chiral limit.
One such multiplet is expected to contain a scalar particle, a pseudoscalar particle, and a fermionic `gluino--glue' particle.
Degeneracy in these channels, in the chiral--continuum limit, was convincingly observed by \refcite{Ali:2019agk}, overcoming numerical challenges that include fermion-line-disconnected contributions to all physical two-point functions, as well as mixing between glueballs and meson-like singlet states.

An additional challenge is carrying out the chiral extrapolations in the presence of an unprotected gaugino mass.
This is done by taking the $m_{\pi}^2 \to 0$ limit for an `adjoint pion' defined in partially quenched chiral perturbation theory~\cite{Munster:2014cja}, which is measured from just the connected part of the correlator for the $\eta'$-like `gluinoball'.
Supersymmetric Ward identities provide an alternative means to determine the chiral limit~\cite{Ali:2018fbq, Ali:2020mvj}.
At a non-zero lattice spacing, any differences between these two determinations can be considered a measure of the supersymmetry-breaking discretization artifacts.
\refcite{Ali:2020mvj} finds that these vanish $\propto$$a^2$, as expected for clover-improved Wilson fermions, supporting the restoration of susy in the chiral continuum limit.

Of course, many other lattice $\cN = 1$ SYM investigations are valuable to carry out in addition to calculations of the gaugino condensate, composite spectrum, and Ward identities.
These include explorations of the finite-temperature phase diagram, which features both a chiral transition related to gaugino condensation as well as a confinement transition related to spontaneous center symmetry breaking.
Refs.~\cite{Bergner:2014saa, Bergner:2019dim} find that these two transitions occur at roughly the same critical temperature, at least for gauge group SU($2$), which was not known a priori.
In addition, Refs.~\cite{Bergner:2014dua, Bergner:2018unx} investigate the phase diagram on $\Rbb^3 \X S^1$ with a small radius for the compactified temporal direction, comparing thermal and periodic boundary conditions (BCs) for the gauginos.
This work finds evidence that periodic BCs allow the confined, chirally broken phase to persist for weak couplings where analytic semi-classical methods~\cite{Poppitz:2012sw} may be reliable.

In these investigations, the gradient flow has helped enable the precise measurement of the gaugino condensate with clover-improved Wilson fermions.
The gradient flow is also widely used to set the scale and extrapolate to the continuum limit in lattice studies of $\cN = 1$ SYM~\cite{Bergner:2014saa, Bergner:2014ska, Ali:2018dnd, Ali:2018fbq, Ali:2019gzj, Ali:2019agk, Ali:2020mvj}, including the recent \refcite{Butti:2022sgy}, which uses twisted Eguchi--Kawai volume reduction to study the theory on a `lattice' that consists of a single site.
Another potential application of the gradient flow is to define renormalized supercurrents and help guide fine-tuning, by constructing a flow that is consistent with supersymmetry in Wess--Zumino gauge~\cite{Hieda:2017sqq, Kasai:2018koz, Kadoh:2018qwg}.
This complements the ongoing use of lattice perturbation theory to analyze these supercurrents, and other operators~\cite{Costa:2020keq, Costa:2021pfu, Bergner:2022wnb}.
Finally, given the progress in algorithms and computing hardware, it is also compelling to continue exploring the use of overlap~\cite{Piemonte:2020wkm} or domain-wall fermions to investigate $\cN = 1$ SYM.

\section{\label{sec:max}Maximal $\cN = 4$ SYM in four dimensions} 
$\cN = 4$ SYM, with $Q = 16$ supercharges in four dimensions, turns out to be another special supersymmetric gauge theory that is promising to consider on the lattice.
This is serendipitous, given the large role that $\cN = 4$ SYM plays in theoretical physics thanks to its many supersymmetries, large SU(4)$_R$ symmetry and conformal symmetry --- with additional simplifications in the large-$N$ planar limit.
Among many other important applications, it is the conformal field theory of the original AdS/CFT holographic duality~\cite{Maldacena:1997re}, and provided early insight into S-duality~\cite{Osborn:1979tq}.
This provides many compelling targets for non-perturbative lattice investigations of $\cN = 4$ SYM to pursue, complementing the many analytic approaches that have already been brought to bear.

Because $\cN = 4$ SYM features six massless real scalar fields in addition to the gauge field and four massless Majorana fermions, a naive lattice discretization would be problematic.
At least $8$ fine-tunings would be required if all the supersymmetries were broken~\cite{Catterall:2014mha}.
Fortunately, $\cN = 4$ SYM is the only $d = 4$ theory with $Q = 2^d$ large enough to apply the (equivalent) twisted and orbifolded constructions introduced in \secref{sec:2d}.
This allows a single `twisted-scalar' supercharge \cQ to be preserved, dramatically improving the situation.

Twisted lattice $\cN = 4$ SYM~\cite{Catterall:2009it, Kaplan:2005ta, Unsal:2006qp, Catterall:2007kn, Damgaard:2008pa} combines the bosonic fields into five-component complexified gauge links $\left\{\cU_a, \cUbar_a\right\}$.
In order for these five links to symmetrically span four dimensions, the discretization of space-time needs to employ the $A_4^*$ lattice, which features a large $S_5$ point-group symmetry.
The twisted $p$-form fermions $\eta$, $\psi_a$ and $\chi_{ab}$ are respectively identified with the sites, links and oriented plaquettes of this $A_4^*$ lattice.
A single fine-tuning of a marginal operator may be required to recover the continuum twisted $\SO{4}_{\text{tw}}$ from the discrete $S_5$ symmetry.
The combination of \cQ and $\SO{4}_{\text{tw}}$ then ensures the restoration of the 15 supersymmetries broken by the lattice discretization~\cite{Catterall:2013roa, Catterall:2014mha, Catterall:2014vga}.
Most numerical calculations so far don't explore this potential fine-tuning, instead fixing the corresponding coefficient to its classical value.

As described in \secref{sec:2d}, this twisting procedure results in $\U{N} = \SU{N} \otimes \U{1}$ gauge invariance, and numerical calculations need to regulate flat directions in both the SU($N$) and $\U{1}$ sectors.
Simple supersymmetry-breaking scalar potentials like those used in lower dimensions (and removed in the course of extrapolating to the continuum limit) only affect the SU($N$) sector.
Regulating the $\U{1}$ sector is more challenging, in large part because the corresponding artifacts appear much more severe.
Initial studies~\cite{Catterall:2014vka} with a second supersymmetry-breaking potential --- involving the determinant of the plaquette as the simplest gauge-invariant quantity sensitive to the $\U{1}$ sector --- exhibited far larger discretization artifacts than an improved action~\cite{Catterall:2015ira} that instead incorporates this plaquette determinant into a $\cQ$-invariant modification of the moduli equations.
More recent work has abandoned the $\U{1}$ gauge invariance entirely~\cite{Catterall:2020lsi}, justified by the decoupling of the $\U{1}$ sector in the continuum limit.
Despite the large $\U{1}$ effects indicating non-decoupling in previous lattice studies, \refcite{Catterall:2020lsi} was able to observe apparently reasonable results out to unprecedentedly strong lattice 't~Hooft couplings $\lalat \leq 30$.
Moving forward, detailed comparisons with the gauge-invariant improved action would be worthwhile to clarify the role of discretization artifacts and gauge invariance.

\begin{figure*}[tbp]
  \centering
  \includegraphics[height=0.34\linewidth]{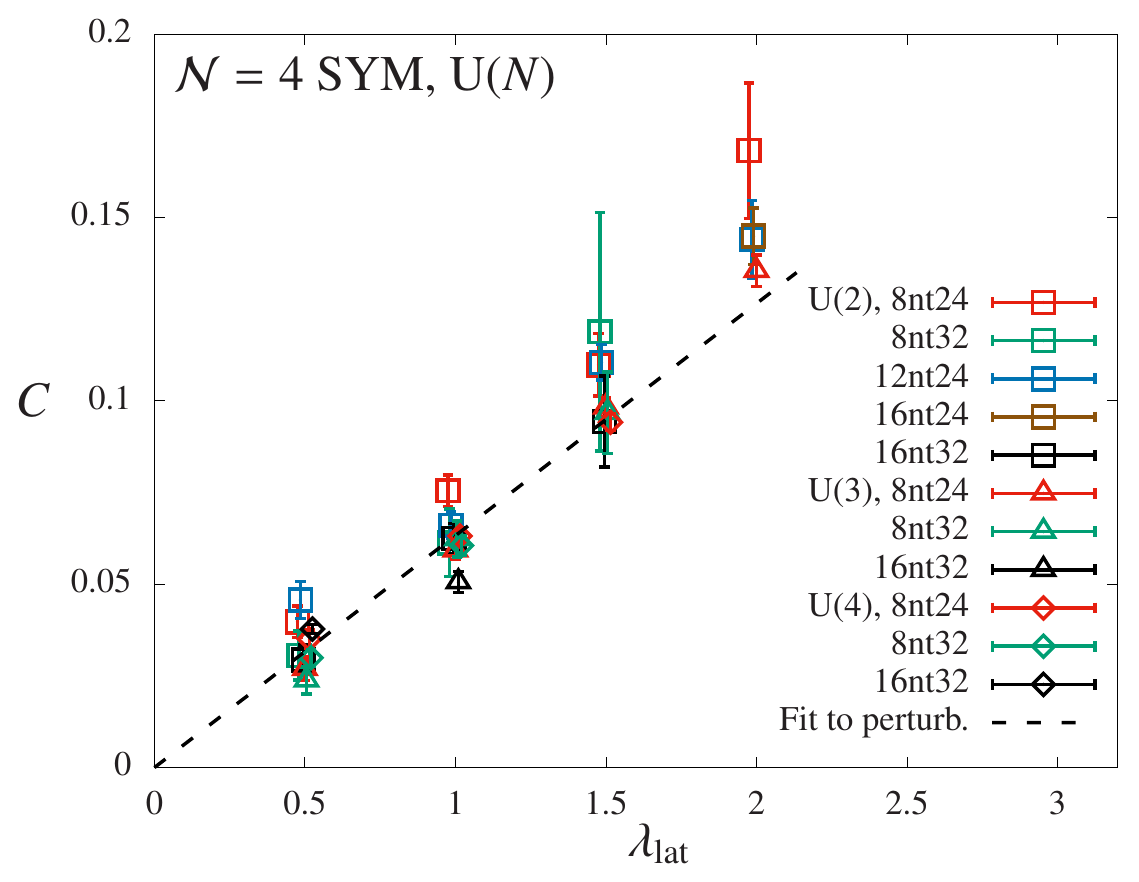}\hfill \includegraphics[height=0.34\linewidth]{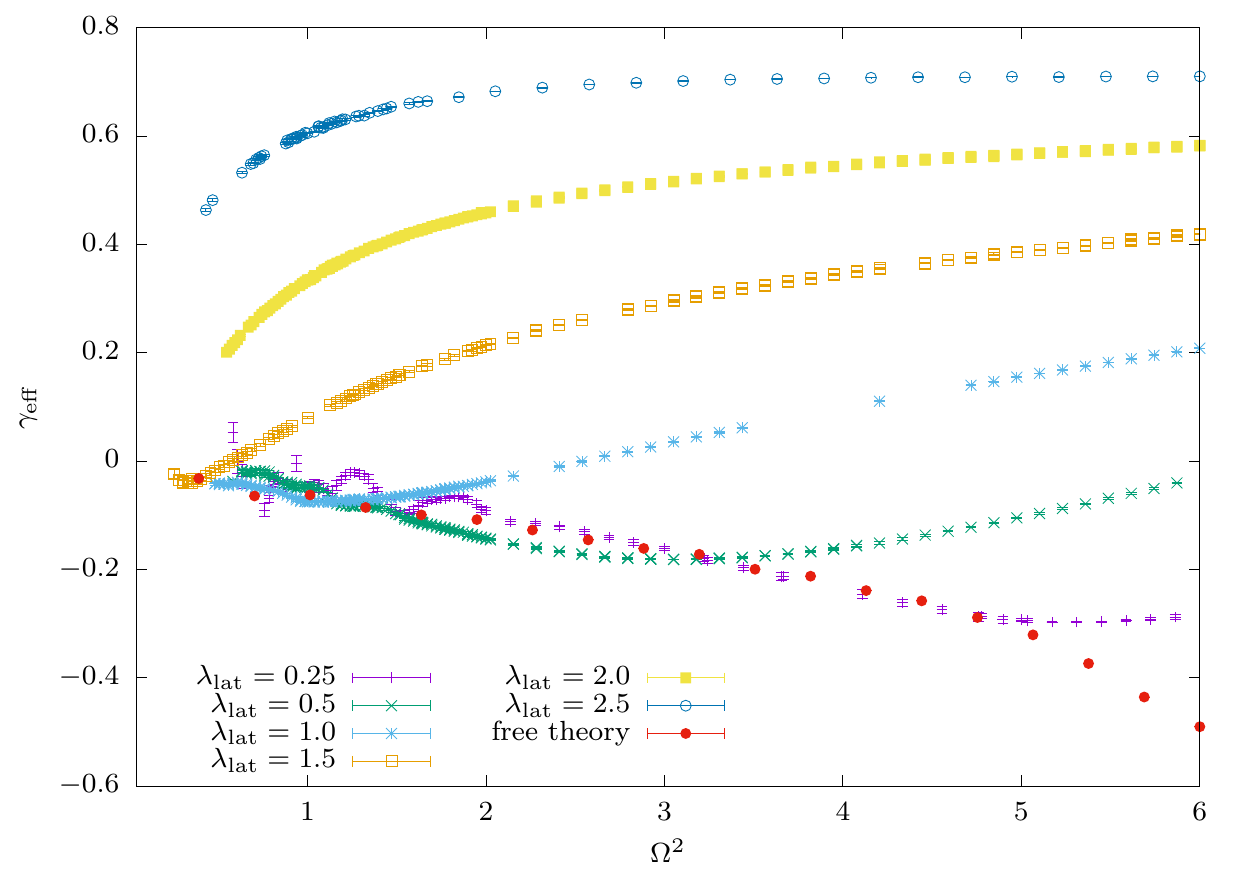}
  \caption{\label{fig:N4SYM}Results from ongoing four-dimensional lattice $\cN = 4$ SYM calculations.  \textbf{Left:} The static potential Coulomb coefficient $C(\lalat)$ is consistent with leading-order perturbation theory (black dashed line) for $\lalat \leq 2$, comparing U($N$) gauge groups with $2 \leq N \leq 4$ and $L^3\X N_t$ lattice volumes with $L \leq 16$ and $N_t \leq 32$.  \textbf{Right:} A scale-dependent effective anomalous dimension $\gaEff(\Om^2)$ approaches the expected $\ga_*(\la) = 0$ in the IR limit $\Om^2 \to 0$.  Considering $16^4$ lattices for gauge group $\U{2}$, discretization artifacts become more significant as the coupling \lalat increases.}
\end{figure*}

The publicly available \texttt{SUSY LATTICE} parallel software package implements the improved action for $\cN = 4$ SYM~\cite{Schaich:2014pda, Bergner:2021ffz_code}, and is currently being used to study a wide range of interesting observables.
For example, the left plot in \fig{fig:N4SYM} considers the static potential $V(r)$, which is correctly seen to be coulombic at all accessible 't~Hooft couplings~\cite{Catterall:2014vka, Catterall:2012yq, Schaich:2016jus, Catterall:2020lsi}.
By incorporating tree-level improvement into the lattice analyses of the static potential~\cite{Schaich:2016jus}, and fitting the resulting data to the Coulomb potential $V(r) = A - C / r$, we obtain the results for the Coulomb coefficient $C(\lalat)$ shown in the plot for several U($N$) gauge groups and lattice volumes.
With $2 \leq N \leq 4$ and $\lalat \leq 2$, the lattice results are consistent with the leading-order perturbative relation $C(\lalat) \propto \lalat$, which isn't surprising given that higher-order perturbative corrections are suppressed by powers of $\frac{\la}{2\pi^2}$~\cite{Pineda:2007kz, Stahlhofen:2012zx, Prausa:2013qva}.
In the strong-coupling planar regime $\la \to \infty$ with $\la \ll N$, there is a famous holographic prediction that $C(\la) \propto \sqrt{\la}$ up to $\cO\left(\frac{1}{\sqrt{\la}}\right)$ corrections~\cite{Rey:1998ik, Maldacena:1998im}, and more general analytic results have been obtained in the $N = \infty$ planar limit~\cite{Gromov:2016rrp}.
Efforts are underway to search for this behavior, by building on \refcite{Catterall:2020lsi} to access stronger couplings.

As a conformal field theory, $\cN = 4$ SYM is characterized by its $\la$-dependent spectrum of scaling dimensions, which are a more challenging target for lattice calculations to predict.
The right plot in \fig{fig:N4SYM} considers the analogue of the mass anomalous dimension, extracted from the eigenmode number of the lattice $\cN = 4$ SYM fermion operator $D$~\cite{Bergner:2021ffz}.
Unlike the Dirac operator of the QCD-like theories where this approach was developed~\cite{Cheng:2013eu, Fodor:2016hke, Bergner:2016hip}, the fermion operator is skew-symmetric, $\Psi^T D \Psi = \chi_{ab} \cD^{(+)}_{[a} \psi_{b]} + \eta \cD^{\dag (-)}_a \psi_a + \frac{1}{2}\eps_{abcde} \chi_{ab} \cD^{\dag (-)}_c \chi_{de}$, and the corresponding anomalous dimension $\ga_*(\la) = 0$ for all couplings.
The eigenmode number $\nu(\Om^2)$ counts the number of $D^{\dag} D$ eigenmodes with eigenvalues $|\la_k|^2 \leq \Om^2$, and scales $\nu(\Om^2) \propto \left(\Om^2\right)^{2 / (1 + \ga_*)}$.
Stochastically constructing and integrating a Chebyshev expansion of the spectral density~\cite{Fodor:2016hke, Bergner:2016hip} produces numerical results for the eigenmode number.
Fitting these results to a power-law within windows $\left[\Om^2, \Om^2 + \ell\right]$ of fixed length $\ell$ provides a scale-dependent effective anomalous dimension $\gaEff(\Om^2)$, which approaches $\ga_*(\la) = 0$ in the IR limit $\Om^2 \to 0$.
Deviations from zero indicate the scale of lattice artifacts related to the breaking of conformality by the finite lattice volume and non-zero lattice spacing, which become more severe at the 't~Hooft coupling \lalat increases.

This experience with a trivially vanishing anomalous dimension aids ongoing investigations of the non-trivial scaling dimension $\De_K(\la) = 2 + \ga_K(\la)$ of the simplest conformal primary operator of $\cN = 4$ SYM, the Konishi operator $\cO_K = \sum_I \Tr{\Phi^I \Phi^I}$, where $\Phi^I$ are the six real scalar fields of the theory.
As for the static potential Coulomb coefficient, there are predictions for the Konishi scaling dimension from weak-coupling perturbation theory~\cite{Fiamberti:2008sh, Bajnok:2008bm, Velizhanin:2008jd}, from holography at strong couplings $\la \to \infty$ with $\la \ll N$~\cite{Gubser:1998bc}, and for all couplings in the $N = \infty$ planar limit~\cite{Gromov:2009zb}.
In addition, due to the conjectured S-duality of the theory, which predicts an invariant spectrum of anomalous dimensions under the interchange $\frac{4\pi N}{\la} \llra \frac{\la}{4\pi N}$, the perturbative results are also relevant in the alternate strong-coupling regime $\la \gg N$~\cite{Beem:2013hha}.
Finally, the superconformal bootstrap program has been applied to analyze the Konishi anomalous dimension, with initial bounds on the maximum value $\ga_K$ could reach across all $\la$~\cite{Beem:2013qxa, Beem:2016wfs} recently being generalized to $\la$-dependent constraints~\cite{Chester:2021aun}.\footnote{Bootstrap results are also available for other superconformal systems including four-dimensional $\cN = 2$ gauge theories and three-dimensional gauge theories --- see Refs.~\cite{Chester:2022sqb, Alday:2021ymb} and references therein.}
Preliminary lattice results with $\lalat \leq 3$ in \refcite{Schaich:2018mmv}, obtained from Monte Carlo renormalization group (MCRG) stability matrix analyses, again appear consistent with perturbation theory.

Numerical lattice analyses of $\cN = 4$ SYM clearly remain in their early stages, with many opportunities for both technical improvements as well as generalizations to other interesting targets.
A great deal of effort is currently focused is on accessing stronger 't~Hooft couplings, both to make more direct contact with holographic predictions and also to investigate the behavior of the system around the S-dual point $\la_{\text{sd}} = 4\pi N$.
We will revisit this issue in \secref{sec:sign}.
Another direction proposed by \refcite{Giedt:2016thz} is to adjust the scalar potential so as to study the theory on the Coulomb branch of the moduli space, where its U($N$) gauge invariance is higgsed to $\U{1}^N$.
In this context S-duality relates the masses of the $\U{1}$-charged elementary `$W$~bosons' and the magnetically charged topological 't~Hooft--Polyakov monopoles~\cite{Osborn:1979tq}, each of which may be accessible from lattice calculations with either C-periodic or twisted BCs, for values of \lalat that have already been studied successfully.
The behavior of lattice $\cN = 4$ SYM at non-zero temperatures will also be interesting to explore.
In particular, there is motivation~\cite{Hanada:2016jok} to study the free energy, with the aim of using non-perturbative lattice calculations to connect the weak-coupling perturbative prediction~\cite{Fotopoulos:1998es} and the strong-coupling holographic calculation~\cite{Gubser:1998bc}, which differ by a famous factor of $\frac{3}{4}$.

\section{\label{sec:future}Challenges for the future} 
While the recent progress in the three areas discussed above is substantial, many compelling directions remain for further research, ranging from improving large-$N$ continuum extrapolations in lower dimensions, to revisiting $\cN = 1$ SYM with Ginsparg--Wilson fermions, and pursuing stronger 't~Hooft couplings in $\cN = 4$ SYM calculations.
There are also many other aspects of supersymmetric gauge theories that have proven more challenging to tackle on the lattice.
In this section we'll conclude this brief review by commenting on two particular challenges --- lattice analyses of superQCD and the possibility of sign problems in supersymmetric lattice systems.

\subsection{Supersymmetric QCD} 
Considering first $\cN = 1$ theories in four dimensions, the generalization from SYM to superQCD involves adding `matter' multiplets --- `quarks' and `squarks', which could in principle transform in any representation of the gauge group.
This generalization would enable investigations of many important phenomena, including (metastable) dynamical supersymmetry breaking, conjectured electric--magnetic dualities and RG flows to known conformal IR fixed points.
Of course, the presence of the scalar squarks implies many more relevant supersymmetry-violating operators, making the necessary fine-tuning far more challenging.
Even exploiting the continuum-like symmetries preserved by overlap or domain-wall fermions, \refcite{Giedt:2009yd} counts $\cO(10)$ operators to be fine-tuned, depending on the gauge group and matter content.
The simplifications offered by the Ginsparg--Wilson relation appear particularly crucial in the context of superQCD.
In particular, Refs.~\cite{Giedt:2009yd, Elliott:2008jp} argue that this good control over the fermions might allow the scalar masses, Yukawas and quartic couplings to be fine-tuned ``offline'' through multicanonical reweighting, significantly reducing computational costs.

While the recent overlap investigation of $\cN = 1$ SYM in \refcite{Piemonte:2020wkm} provides a starting point for generalization to superQCD, most recent lattice explorations use Wilson fermions and are confronted with the full fine-tuning challenge.
One tactic is to proceed by using lattice perturbation theory to guide numerical calculations~\cite{Costa:2017rht, Costa:2018mvb, Wellegehausen:2018opt, Costa:2020keq, Costa:2021pfu, Bergner:2022wnb}.
There is also an initial investigation of a superQCD gradient flow that is consistent with supersymmetry in Wess--Zumino gauge~\cite{Kadoh:2019flv}.
Another approach is to omit the scalar fields at first, and initially study gauge--fermion theories including both adjoint gauginos and fundamental quarks~\cite{Bergner:2020mwl, Bergner:2021ivi}.
A similar strategy is also being applied to $\cN = 2$ SYM, using overlap fermions~\cite{Bergner:2022hoo}.
These scalar-less studies also provide useful connections to investigations of near-conformal composite Higgs models, recently reviewed by Refs.~\cite{Witzel:2019jbe, Drach:2020qpj}.
All of these studies in four dimensions remain in their early stages.

As an alternative, following the logic of \secref{sec:lowd}, it can prove advantageous to investigate superQCD in the simpler setting of fewer than four space-time dimensions.
Going all the way down to 0+1~dimensions, for example, Refs.~\cite{Filev:2015cmz, Asano:2016xsf, Asano:2016kxo} consider the Berkooz--Douglas matrix model~\cite{Berkooz:1996is}, which adds $N_f$ fundamental multiplets to $Q = 16$ SYM QM in such a way as to preserve half of the supercharges in the continuum.
In $d = 2$ and $d = 3$ dimensions, most effort so far has focused on constructing clever lattice formulations of superQCD~\cite{Matsuura:2008cfa, Sugino:2008yp, Kikukawa:2008xw, Kadoh:2009yf, Joseph:2013jya, Joseph:2013bra, Joseph:2014bwa}.

Existing numerical calculations~\cite{Catterall:2015tta}, and work in progress~\cite{Sherletov:2022rnl}, employ a quiver construction of $Q = 4$ superQCD in two dimensions~\cite{Matsuura:2008cfa, Sugino:2008yp}, based on the twisted formulation introduced in \secref{sec:2d}.
Starting from $Q = 8$ SYM in three dimensions, the system is reduced to have only two slices in the third direction.
The twisted formulation is then generalized to have different gauge groups $\U{N}$ and $\U{F}$ on each of these two-dimensional slices.
Decoupling the $\U{F}$ slice then produces a two-dimensional U($N$) theory with $F$ massless fundamental matter multiplets and $Q = 4$ supersymmetries, one of which is preserved at non-zero lattice spacing.
This same approach can be applied to construct $Q = 8$ superQCD in two and three dimensions~\cite{Joseph:2013jya, Joseph:2013bra}, and may be generalizable to higher representations~\cite{Joseph:2014bwa}.
The main numerical result so far has been to compare $\U{2}$ superQCD with $F = 3$ against $\U{3}$ superQCD with $F = 2$, observing dynamical supersymmetry breaking when $N > F$ and confirming that the resulting goldstino is consistent with masslessness in the infinite-volume limit~\cite{Catterall:2015tta}.

\subsection{\label{sec:sign}Sign problems} 
Another challenge confronting some supersymmetric lattice field theories is the possibility of sign problems, at least in certain regimes.
\refcite{deForcrand:2010ys} provides a brief general introduction to sign problems.
Here we will return to focusing on SYM, which involves only Majorana fermions, and hence the pfaffian of the fermion operator $D$, which can fluctuate in sign even when the determinant would be positive.
Separating a generic complex pfaffian into its magnitude and phase, $\pf D = |\pf D| e^{i\al}$, the results presented in \secref{sec:max} all come from `phase-quenched' RHMC calculations that use only $|\pf D|$ to carry out importance sampling.
The resulting phase-quenched observables $\vev{\cO}_{\text{pq}}$ then need to be reweighted, $\vev{\cO} = \vev{\cO}_{\text{pq}} / \vev{e^{i\al}}_{\text{pq}}$, with a sign problem appearing when $\vev{e^{i\al}}_{\text{pq}} = Z / Z_{\text{pq}}$ vanishes within statistical uncertainties.
In particular, in lattice calculations employing fully periodic BCs, the partition function $Z$ is the Witten index and must vanish for any theory that can exhibit spontaneous supersymmetry breaking~\cite{Witten:1982df}, implying a severe sign problem.

For lattice $\cN = 1$ SYM, clover-improved Wilson fermions produce a real pfaffian whose sign can be computed efficiently~\cite{Bergner:2011zp}.
\refcite{Ali:2018dnd} reports that $\vev{e^{i\al}}_{\text{pq}} \approx 1$ in these studies, improving as the lattice spacing decreases.
The twisted-mass approach to $\cN = 1$ SYM allows the pfaffian to be complex.
While \refcite{Steinhauser:2020zth} finds $\vev{e^{i\al}}_{\text{pq}} \approx \cos(\al) > 0.965$ for $16^3\X 32$ lattices (extrapolating from smaller volumes), it also reports that $\vev{e^{i\al}}_{\text{pq}}$ decreases exponentially in the lattice volume $V$, as expected~\cite{deForcrand:2010ys}.
This work has to extrapolate from smaller volumes because it directly evaluates the pfaffian, which can be extremely expensive, with computational costs scaling $\propto$$N_{\Psi}^3$, where $N_{\Psi} \propto N^2 V$ is the number of fermionic degrees of freedom in the lattice system.
Such direct evaluations are more common in lower dimensions, where all works so far observe well-behaved $\vev{e^{i\al}}_{\text{pq}} \to 1$ in the fixed-$\That$ continuum limit~\cite{Catterall:2009xn, Filev:2015hia, Hanada:2010qg, Kamata:2016xmu, Catterall:2011aa, August:2018esp, Catterall:2017xox, Schaich:2022duk, Sherletov:2022rnl}.
Lower-dimensional systems can also be used as testbeds for different algorithmic approaches, such as the complex Langevin method used to explore spontaneous supersymmetry breaking in \refcite{Joseph:2020gdh}.

\begin{figure*}[tbp]
  \centering
  \includegraphics[width=0.48\linewidth]{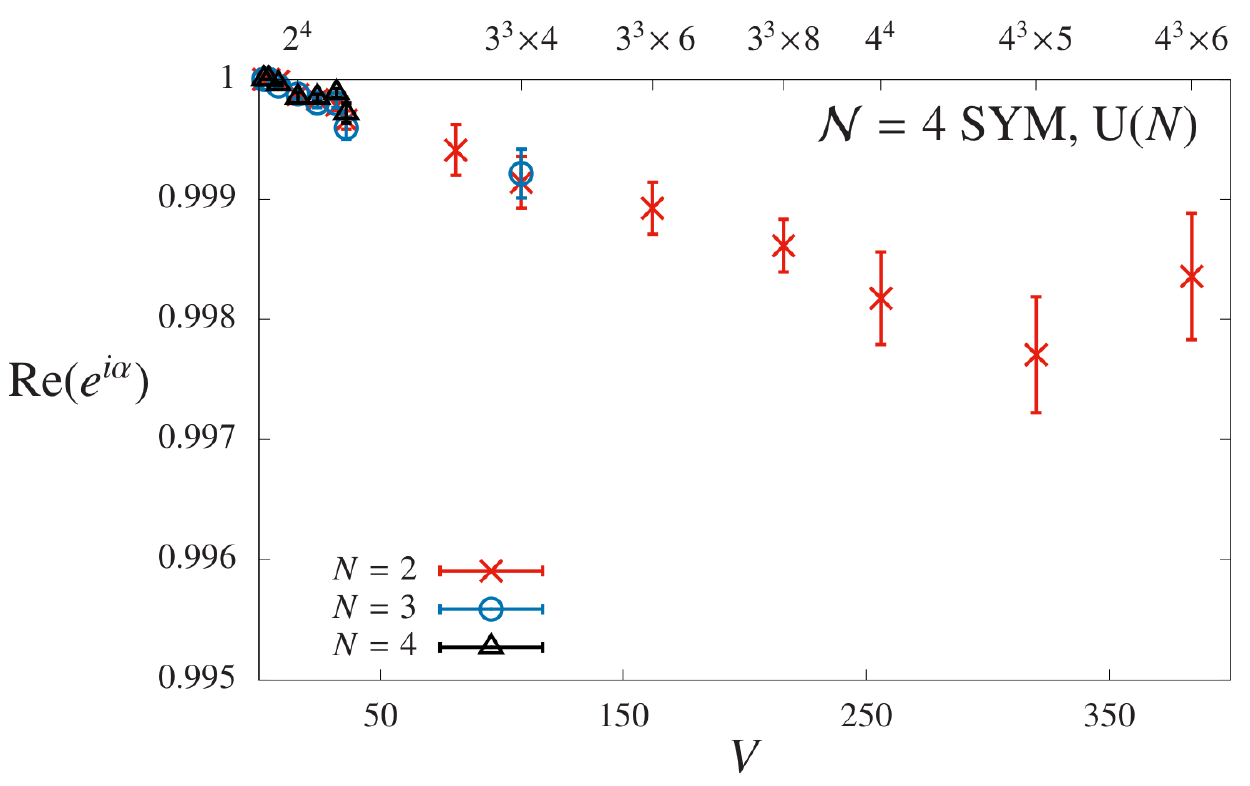}\hfill \includegraphics[width=0.48\linewidth]{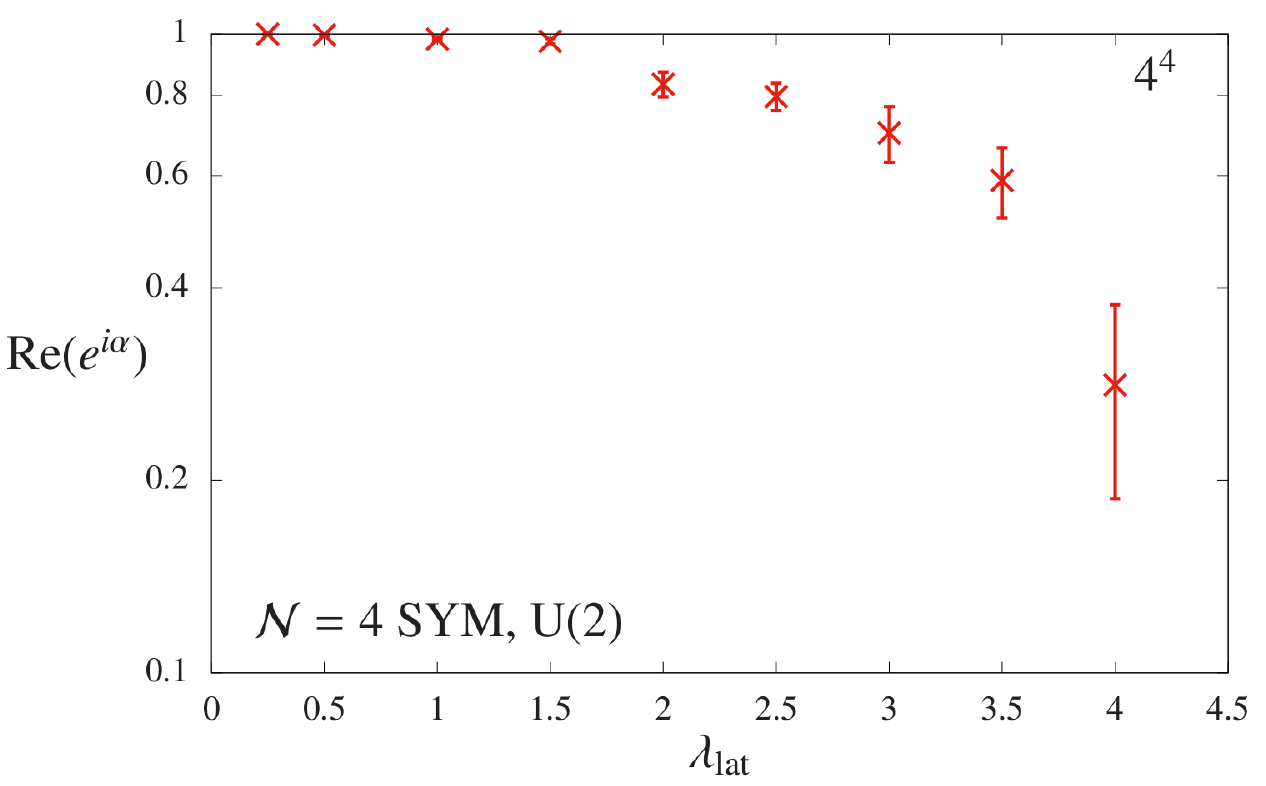}
  \caption{\label{fig:N4phase}Results for the pfaffian phase $\vev{\mbox{Re} \left(e^{i\al}\right)}_{\text{pq}} \approx \vev{e^{i\al}}_{\text{pq}}$ from four-dimensional lattice $\cN = 4$ SYM calculations using the gauge-invariant improved action~\cite{Catterall:2014vga, Schaich:2015daa}.  \textbf{Left:} For a fixed 't~Hooft coupling $\lalat = 0.5$, only per-mille-level fluctuations are observed for U($N$) gauge groups with $N = 2$, 3 and 4, up to the largest accessible volumes.  \textbf{Right:} For a fixed $4^4$ lattice volume, the $\U{2}$ phase fluctuations increase significantly for stronger couplings, which obstructs studies of $\lalat \gsim 4$ with this lattice action.}
\end{figure*}

Turning to lattice $\cN = 4$ SYM in four dimensions, \fig{fig:N4phase} presents results for the pfaffian phase using the gauge-invariant improved action~\cite{Catterall:2014vga, Schaich:2015daa}.
Despite implementing a parallelized pfaffian computation in \texttt{SUSY LATTICE}, only small $N$ and small lattice volumes are computationally accessible, with each $\U{2}$ $4^4$ pfaffian measurement requiring approximately $50$~hours on $16$~cores.
In the left plot, only small per-mille-level phase fluctuations are observed on all accessible volumes with fixed 't~Hooft coupling $\lalat = 0.5$.
In particular, the expected exponential suppression of $\vev{e^{i\al}}_{\text{pq}}$ with the lattice volume is not visible, which is encouraging if not yet fully understood.
However, the right plot shows that $\vev{e^{i\al}}_{\text{pq}}$ decreases rapidly as the 't~Hooft coupling increases, obstructing studies of $\lalat \gsim 4$ with this lattice action.
For gauge group $\U{2}$, \refcite{Catterall:2020lsi} presents mixed-action evidence that sacrificing gauge invariance may improve control over pfaffian phase fluctuations at much stronger couplings $\lalat \sim \cO(10)$, raising the possibility of more directly probing holography and S-duality.

\subsection{Final remarks} 
Non-perturbative lattice investigations of supersymmetric QFTs are important and challenging, making this an area that should attract even more attention in the near future.
It is encouraging that there has been so much recent progress in lattice studies of four-dimensional $\cN = 1$ SYM and $\cN = 4$ SYM, along with their dimensional reductions to $d < 4$, where the consequences of supersymmetry breaking due to the discrete lattice space-time can be kept under control.
There have also been advances in other areas of lattice supersymmetry that this brief review omits, in particular lattice studies of theories without gauge invariance, such as Wess--Zumino models and sigma models~\cite{Wozar:2011gu, Steinhauer:2014yaa, Baumgartner:2014nka, Baumgartner:2015qba, Baumgartner:2015zna, Aoki:2017iwi, Kadoh:2018hqq, Kadoh:2018ele, Kadoh:2018ivg, Kadoh:2019glu, Joseph:2020gdh, Dhindsa:2020ovr, Culver:2021rxo, Feng:2022nyi}, as well as lattice investigations of the Green--Schwarz superstring worldsheet sigma model~\cite{Bianchi:2016cyv, Bianchi:2019ygz, Forini:2021keh, Bliard:2022kne}.
While supersymmetric QCD and sign problems present challenges that may be difficult to overcome, the overall prospects of lattice supersymmetry are bright, with many compelling opportunities for future work.

\section*{Acknowledgments} 
I thank the International Centre for Theoretical Sciences in Bengaluru for hosting the 2021 online program ``Nonperturbative and Numerical Approaches to Quantum Gravity, String Theory and Holography'' (ICTS/numstrings2021/1), where I gave a pedagogical presentation that provided inspiration for this short review.
My understanding of lattice supersymmetry has benefited greatly from past and present collaborations with Georg Bergner, Simon Catterall, Chris Culver, Poul Damgaard, Tom DeGrand, Navdeep Singh Dhindsa, Joel Giedt, Raghav Jha, Anosh Joseph, Angel Sherletov and Toby Wiseman.
I am supported by UK Research and Innovation Future Leader Fellowship {MR/S015418/1} and STFC grant {ST/T000988/1}. \\[8 pt]

\noindent \textbf{Data Availability Statement:} No new data are generated or analyzed for this review.
See the referenced references for information about data associated with the work being reviewed.

\raggedright
\bibliography{lattice_susy}
\end{document}